\documentclass[aps,prl,showpacs,twocolumn,superscriptaddress]{revtex4}
\usepackage{graphicx}
\usepackage{bm}
\usepackage{amsmath}

\def\be{\begin{equation}}       \def\ee{\end{equation}}
\def\bea{\begin{eqnarray}}      \def\eea{\end{eqnarray}}

\begin{document}

\title{Spin Hall Insulator}
\author{Shuichi Murakami}
\email[Electronic address: ]{murakami@appi.t.u-tokyo.ac.jp}
\affiliation{Department of Applied Physics, University of Tokyo,
Hongo, Bunkyo-ku, Tokyo 113-8656, Japan}
\author{Naoto Nagaosa}
\affiliation{Department of Applied Physics, University of Tokyo,
Hongo, Bunkyo-ku, Tokyo 113-8656, Japan} \affiliation{CERC, AIST
Tsukuba Central 4, Tsukuba 305-8562, Japan} \affiliation{CREST,
Japan Science and Technology Agency (JST)}
\author{Shou-Cheng Zhang}
\affiliation{Department of Physics, McCullough Building, Stanford
University, Stanford CA 94305-4045}

\begin{abstract}
Recent theories predict dissipationless spin current induced by an
electric field in doped semiconductors. Nevertheless, the charge
current is still dissipative in these systems. In this work, we
theoretically predict the dissipationless spin Hall effect,
without any accompanying charge current, in some classes of band
insulators, including zero-gap semiconductors such as HgTe and
narrow-gap semiconductors such as PbTe. This effect is similar
to the quantum Hall effect in that all the states below the gap
contribute and there occurs no dissipation. However the spin Hall
conductance is not quantized even in two dimensions. This is the
first example of a nontrivial topological structure in a band
insulator without any magnetic field.
\end{abstract}

\maketitle

{\it Introduction.}--- Generation of the spin current by an
external electric field
\cite{murakami2003,sinova2003} has attracted recent interests. It
has been proposed theoretically that hole-doped semiconductors
such as p-GaAs and p-Ge show this effect due to the intrinsic
spin-orbit interaction in the Luttinger model\cite{murakami2003}.
The n-GaAs with the inversion symmetry breaking is another
candidate for this effect\cite{sinova2003}, although some
questions regarding disorder effects still remain \cite{inoue2004, 
murakami2004}. 
In both cases, the spin current is generated not by
the displacement of the electron distribution function, but by the
anomalous velocity due to the Berry curvature of the Bloch states.
In this sense, the spin current does not accompany dissipation,
and it is similar to both the quantum Hall effect (QHE) and also
the intrinsic anomalous Hall effect (AHE). However, even though
the spin current itself does not cause Joule heating, the applied
electric field generates a charge current because of nonzero
resistivity. Hence, strictly speaking, it is not
dissipationless as a whole system. When the carrier density is
reduced, the charge conductivity decreases faster than the spin
conductivity. Therefore, both for low power logic device
applications and for theoretical interest, we are lead to the
question of whether the spin Hall effect can exist in a band
insulator without any charge current. In conventional semiconductors
such as GaAs and Si without doping, the spin Hall effect is 
zero; they are inert both for the spin and the charge responses, 
as assumed in conventional band theories.

In contrast to the conventional semiconductors, 
we propose in this Letter two classes of band insulators showing
finite spin Hall conductivity without any charge conductivity. One
class is (distorted) zero-gap semiconductors such as HgTe,
HgSe, $\beta$-HgS and $\alpha$-Sn. In these systems, in contrast
to conventional semiconductors, the heavy-hole (HH) and light-hole
(LH) bands have the opposite signs of the mass, with
only the former being occupied at zero doping, and the finite spin
Hall effect is predicted. Furthermore, by introducing a uniaxial
strain, a gap opens between the HH and LH bands without
destroying the spin Hall effect. The other class is narrow-gap
semiconductors such as PbTe, PbSe and PbS, where  even without
doping the spin Hall conductance is nonzero. Thus these two
classes of materials without doping have a gap and the spin Hall
current accompanies no dissipation, similar to the QHE. Unlike the 
QHE, however, these band insulators have severeal new aspects;
(i) 
the spin Hall conductance is not quantized and depends on
parameters characterizing the band structure, 
(ii) there is no magnetic field, and the system is three-dimensional,
and 
(iii) the effect is protected by the large band gap and is robust even 
at room temperature.
From these novel aspects, this spin Hall effect 
realizes new state of matter in the simple band insulators.

These novel features originate from the fact that the gap in these 
materials arises due to the spin-orbit coupling, which 
causes the spin Hall effect.
The spin-orbit coupling gives rise to a splitting of bands into
multiplets of the total angular momentum
$\mathbf{J}=\mathbf{L}+\mathbf{S}$. If all the bands in the
same $J$ multiplet are filled, they do not contribute to spin Hall
conductivity. Only when the fillings of the bands in the same
$J$ multiplet are different, the spin Hall conductivity can be
nonzero. Nonzero spin Hall
effect does not require breaking of any symmetry such as inversion
or time-reversal. Thus the spin Hall effect should be a common
effect, while its magnitude may vary from material to material.
In particular, the spin Hall effect 
is nonzero in nonmagnetic band insulators
such as zero-gap and narrow-gap semiconductors, as explained below.
This is distinct from the spin current in a 
spin system considered by Meier and
Loss \cite{meier2003}.

To calculate the spin Hall conductivity $\sigma_{s}$, we follow
the method developed by the authors \cite{murakami2003c}.
We restrict ourselves to systems with
time-reversal and inversion symmetries;
therefore, all the states form Kramers doublets. By picking up two doubly
degenerate bands near the Fermi energy, the Hamiltonian is written as a linear
combination of 4$\times$4 matrices $\Gamma_{a} (a=1,2,3,4,5)$,
which form the Clifford algebra
$\{\Gamma_{a},\Gamma_{b}\}=2\delta_{ab}$. 
Following \cite{murakami2003c}, we
write the Hamiltonian \cite{note-Hamiltonian} as
\begin{equation}H=\epsilon(\mathbf{k})
+\sum_{a=1}^{5}d_{a}(\mathbf{k})
\Gamma_{a},\label{Hamiltonian}
\end{equation}
where $\epsilon(\mathbf{k})$ and $d_{a}(\mathbf{k})$ are even functions
of $\mathbf{k}$. Its eigenvalues are given by $\epsilon_{\pm}(\mathbf{k})
=\epsilon(\mathbf{k})\pm d(\mathbf{k})$, where $d=|\mathbf{d}|$.

{\it Zero-gap semiconductors.}--- As a first example, we consider
zero-gap semiconductors with diamond or zincblende structures.
Examples are $\alpha$-Sn for the former and HgTe, HgSe, and
$\beta$-HgS for the latter. In the zincblende structure, the
inversion symmetry breaking is small and can be neglected. 
In 
experiments \cite{exp} and first-principle calculations \cite{first} for
HgSe, however,
there exists some
controversy on their band structure near the $\Gamma$ point,
whether the gap is zero or finite.
Here we consider the zero-gap case, while the
following discussions and estimates are not essentially affected
by minute difference of band structures at the $\Gamma$ point.

Before going to the zero-gap semiconductors,
let us review the calculation of the spin Hall effect
for p-type conventional semiconductors with diamond or
zincblende structures \cite{murakami2003c}.
In the Hamiltonian Eq.~(\ref{Hamiltonian}), the $\Gamma_{a}$
matrices 
are the five traceless matrices quadratic in spin-3/2
matrices \cite{luttinger1956,avron1988}. 
The explicit forms of $\Gamma_{a}$ are
given in \cite{murakami2003c}. 
In \cite{murakami2003c}, to overcome the difficulty in defining a
spin current, which comes from spin non-conservation 
due to the spin-orbit interaction, the spin $\mathbf{S}$ has 
been separated into a conserved part
$\mathbf{S}^{(c)}$ and a non-conserved part $\mathbf{S}^{(n)}$:
$\mathbf{S}=\mathbf{S}^{(c)}+ \mathbf{S}^{(n)}$. The conserved
part $\mathbf{S}^{(c)}$ consists of intraband matrix elements of
the spin. From this conserved spin $\mathbf{S}^{(c)}$ we can
uniquely define a conserved spin current from the Noether's
theorem. The spin Hall conductivity calculated by the Kubo formula
is given as
\begin{equation}
\sigma_{ij\text{(c)}}^{l} =\frac{4}{3V}\sum_{\mathbf{k}}
(n_{L}(\mathbf{k})-n_{H}(\mathbf{k})) \eta_{ab}^{l}G_{ij}^{ab},
\label{sigmaijl}\end{equation} where $n_{L}(\mathbf{k})$,
$n_{H}(\mathbf{k})$ are the Fermi distributions of 
holes in the LH and the HH bands. We take the hole picture
in this model, where the LH and HH bands have positive energies.
The geometric tensor $G_{ij}^{ab}$  is calculated 
as
\begin{equation} G^{ab}_{ij}=\frac{1}{4d^{3}}
\epsilon_{abcde}d_{c}\frac{\partial d_{d}}{\partial k_{i}
}\frac{\partial d_{e}}{\partial k_{j}}, \label{Gij}\end{equation}
where $\epsilon_{abcde}$ is the totally antisymmetric tensor with 
$\epsilon_{12345}=1$.
Each element of the tensor $\eta_{ab}^{l}$ is a constant,
and its expression is given in
\cite{murakami2003c}. For cubic semiconductors we can write
$\sigma_{ij\text{(c)}}^{l}=\epsilon_{ijl}\sigma_{s}$, where
$\sigma_{s}$ is a constant.

\begin{figure}[h]
\includegraphics[scale=0.48]{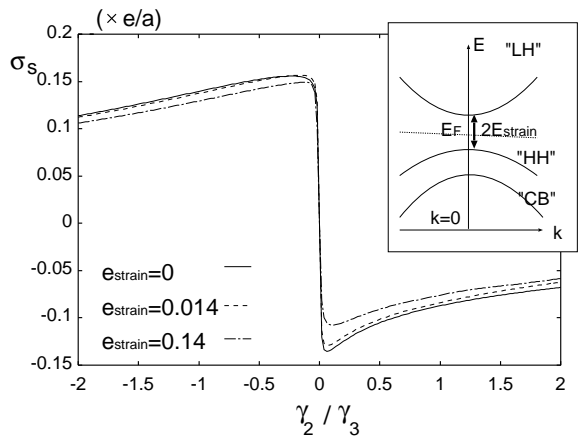}
\caption{
Spin Hall conductivity $\sigma_{s}$ as a function of
$\gamma_{2}/\gamma_{3}$, calculated from the tight-binding
Hamiltonian modelling a zero-gap semiconductor without doping.
$e_{\text{strain}}=ma^{2}E_{\text{strain}}/\gamma_{3}$ 
is a dimensionless parameter representing 
the uniaxial compressive strain, where
$2E_{\text{strain}}$ is an energy splitting at $
\mathbf{k}=\mathbf{0}$ due to the strain.
$e_{\text{strain}}\sim 0.014$ corresponds to the value of 
$\alpha$-Sn in a compressive stress $3.18\times 10^9\text{dyn/cm}^{2}$ 
\cite{roman1972}.
Inset: a schematic picture of the band structure for the zero-gap
semiconductors with uniaxial strain. 
The labels LH, HH and CB corresponds to the light-hole, heavy-hole, and
the conduction bands in the conventional semiconductors with cubic symmetry,
respectively. 
$E_{F}$ is the Fermi energy.}
\label{fig-bulk}\end{figure}

The zero-gap semiconductors have ``inverted'' band structure,
compared with conventional semiconductors, 
as shown in the inset of Fig.~\ref{fig-bulk}. The energy
of the original conduction band becomes lower than the Fermi
energy $E_{F}$. At the same time, the LH band moves up to become a
conduction band, while the HH band remains a top of the valence
band. The ``LH'' and ``HH'' bands touch at the $\Gamma$ point
($\mathbf{k}=0$). Although the band structure is largely different 
from conventional ones, its
gauge-field structure giving rise to the intrinsic 
spin Hall effect remains the same, and we can still use
Eq.~(\ref{sigmaijl}). In
Eq.~(\ref{sigmaijl}), the summand is proportional to
$n_{L}(\mathbf{k})-n_{H}(\mathbf{k})$. Therefore, in conventional
semiconductors, hole-doping is required for nonzero spin Hall
conductivity $\sigma_{s}$; in remarkable contrast, 
$\sigma_{s}$ is nonzero in the zero-gap
semiconductors even without doping, because $n_{L}(\mathbf{k})=1$,
and $n_{H}(\mathbf{k})=0$ \cite{note-nLnH}.

In conventional semiconductors
$n_{L}-n_{H}$ is nonzero only near the $\Gamma$ point, and to
calculate $\sigma_{s}$ we could use the Luttinger Hamiltonian
valid only for small $\mathbf{k}$
\cite{murakami2003,murakami2003c}. On the contrary, to calculate
the spin Hall conductivity for the zero-gap semiconductors, we
need a Hamiltonian valid for all $\mathbf{k}$. Hence,
we shall construct a
tight-binding Hamiltonian. 
The primitive vectors
are $\mathbf{a}_{1}=\frac{1}{2}(0,a,a)$,
$\mathbf{a}_{2}=\frac{1}{2}(a,0,a)$, and
$\mathbf{a}_{3}=\frac{1}{2}(a,a,0)$, where $a$ is a lattice
constant. 
Necessary conditions for the Hamiltonian are (a) cubic
symmetry,
(b) inversion symmetry, and 
(c) time-reversal symmetry. Under these conditions,
the simplest model which reproduces the Luttinger Hamiltonian 
near the $\Gamma$ point is written as
\begin{eqnarray*}
&&d_{1}=\sqrt{3}\gamma_{3}C
 [\cos
{\theta}_{1}-\cos({\theta}_{2}-{\theta}_{3})],\\
&&d_{2}=\sqrt{3}\gamma_{3}C[\cos
{\theta}_{2}-\cos({\theta}_{3}-{\theta}_{1})],\\
&&d_{3}=\sqrt{3}\gamma_{3}C [\cos
{\theta}_{3}-\cos({\theta}_{1}-{\theta}_{2})],\\
&&d_{4}=\sqrt{3}\gamma_{2}C
[\cos \theta_{2}-\cos \theta_{1}\\
&&\ \ \ +\cos(\theta_{3}-\theta_{1})
-\cos(\theta_{2}-\theta_{3})],\\
&&d_{5}=\gamma_{2}C [\cos
\theta_{1}+\cos\theta_{2}
-2\cos({\theta}_{1}-{\theta}_{2})\\
&&\ \ \ +\cos(\theta_{3}-\theta_{1})+\cos
(\theta_{2}-\theta_{3}) -2\cos
{\theta}_{3})],
\end{eqnarray*}
where $C=2/(a^{2}m)$, $m$ is the electron mass, 
and $\theta_{j}=\mathbf{k}\cdot\mathbf{a}_{j}$ ($j=1,2,3$).
The constants $\gamma_{2}$ and
$\gamma_{3}$ correspond to the Luttinger parameters
\cite{luttinger1956}.
In the real-space representation, the spin-dependent part of the 
Hamiltonian involves only the nearest neighbor hopping: 
\begin{eqnarray}
&& H_{\text{spin-dep.}}=\frac{1}{a^{2}m} \sum_{t_{y},t_{z}=\pm 1,\mathbf{x}}
c^{\dagger}_{\mathbf{x}+\frac{a}{2}(0,t_{y},t_{z}),\alpha}
c_{\mathbf{x}\beta}\nonumber \\
&& \ \ \ \
\cdot(-2\gamma_{2}S^{x2}+\gamma_{3}t_{y}t_{z}(S^{y}S^{z}+
S^{z}S^{x}))_{\alpha\beta}\nonumber \\
&&+(\mbox{two cyclic permutations of $x$,$y$,$z$}).
\end{eqnarray}
Although this Hamiltonian is  a simplified one, we
expect that it correctly captures basic physics and an order estimate of the
spin Hall effect.

It is a straightforward task to 
calculate the spin Hall
conductivity by substituting the $\mathbf{d}$ vector into Eqs.\ 
(\ref{sigmaijl}) and (\ref{Gij}). In
Fig.~\ref{fig-bulk} is shown the value of $\sigma_{s}$ as a
function of $\gamma_{2}/\gamma_{3}$. Nominally
$\gamma_{2}/\gamma_{3}\sim 1$ for zero-gap semiconductors, and we
get $\sigma_{s}\sim -0.1 \frac{e}{a}$.
We note that the absolute value of the spin Hall conductivity $\sigma_s$
increases as $\gamma_{2}/\gamma_{3}$ decreases, while in
hole-doped conventional ``uninverted'' semiconductors $\sigma_s$ 
is maximum around $\gamma_2/\gamma_3\sim 1$ 
\cite{bernevig2003}. 

In such case without doping, the longitudinal
conductivity $\sigma\sim 0$ and the system is an insulator, 
whereas the spin Hall conductivity is nonzero. In a 
three-dimensional sample,
the gap is zero, and finite
temperature easily provides the sample with n- and p-type
carriers, producing the longitudinal (charge) conductivity.
In order to circumvent
this temperature effect and to keep the system insulating, it is
desirable to make a finite gap at the $\Gamma$ point. 
Because the degeneracy of the valence and conduction bands at 
the $\Gamma$ point originates from the cubic symmetry, it is 
lifted by lowering the crystal symmetry, for example, by a
uniaxial strain. This is indeed experimentally observed in 
$\alpha$-Sn, where the compressive uniaxial 
stress opens a gap at the $\Gamma$ point  \cite{roman1972}. 
One can calculate an effect of a strain on the band structure
by the method proposed by Pikus and Bir \cite{pikus1959}.
For simplicity, we focus on a uniaxial strain along the $z$-axis.
Within the present framework with $\Gamma$ matrices,
one can incorporate this uniaxial strain as
an additional term $E_{\text{strain}}\Gamma_{5}$ in the Hamiltonian
\cite{pikus1959}. We calculated the spin Hall conductivity
for various values of a dimensionless constant $e_{\text{strain}}
=ma^{2}E_{\text{strain}}/\gamma_{3}$, and the result is
shown in Fig.\ \ref{fig-bulk}.
For $\alpha$-Sn, a compressive strain of $3.18\times 10^9\text{ dyn/cm}^2$ 
induces a energy splitting
of $2E_{\text{strain}}\sim 44.2 \text{meV}$ at $\mathbf{k}=\mathbf{0}$.
Meanwhile, this splitting corresponds to
$e_{\text{strain}}\sim 0.014$, considerably smaller than unity. 
Hence, as seen from Fig.\ \ref{fig-bulk}, the splitting of 
this size or even larger splitting does not affect much the value of 
$\sigma_s$.
Thus, by opening a gap at the $\Gamma$ point by a uniaxial 
strain, the spin Hall remains nonzero, while the charge conductivity is
suppressed. We name it a spin Hall insulator, though the term
``Hall insulator'' have been used in a different context;
the Hall insulator \cite{zhang1992} refers to an insulator where
both $\sigma_{xy}$ and $\sigma_{xx}$ vanish while $R_{xy}$ is
finite.

{\it Narrow-gap semiconductors.}--- Another example of the spin
Hall effect in band insulators is the narrow-gap semiconductors such as
PbTe, PbSe, and PbS \cite{fradkin}. 
The crystal has the rocksalt structure, with
the primitive vectors $\mathbf{a}_{i}$ $(i=1,2,3)$ given above.
The direct gap is
formed at the four equivalent $L$ points
$\mathbf{p}_{1}=\frac{\pi}{a}(1,1,1)$,
$\mathbf{p}_{2}=\frac{\pi}{a}(1,-1,-1)$,
$\mathbf{p}_{3}=\frac{\pi}{a}(-1,1,-1)$ and
$\mathbf{p}_{4}=\frac{\pi}{a}(-1,-1,1)$,
and is of the order of 0.15-0.3 eV.
The valence and the
conduction bands both form Kramers doublets. Near
these $L$ points, the Hamiltonian is given by
\cite{volkov1983,tchernyshyov2000}
\begin{equation}
H=v\mathbf{k}\cdot\hat{p}\tau_{1}+\lambda v \mathbf{k}\cdot
(\hat{p}\times\bm{\sigma})\tau_{2} +Mv^{2}\tau_{3}.
\label{L-point}\end{equation}
Here $\mathbf{k}$ is a wavevector measured
from $\mathbf{p}_{i}$, and
$\hat{p}_{i}=\mathbf{p}_{i}/|\mathbf{p}_{i}|$. $\tau_{j}$ and
$\sigma_{j}$ are the Pauli matrices, corresponding to the orbital
and the spin, respectively.

\begin{figure}[h]
\includegraphics[scale=0.35]{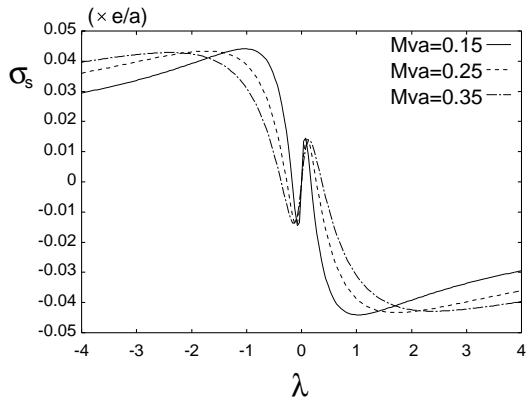}
\caption{Spin Hall conductivity $\sigma_{s}$ as a function of
$\lambda$ for various $m=Mva$, calculated from the tight-binding
Hamiltonian modelling a narrow-gap semiconductor without doping.}
\label{fig-narrow}\end{figure}

 As the second term in the r.h.s. of 
Eq.~(\ref{L-point}) resembles the Rashba Hamiltonian, we expect a
nonzero spin Hall conductivity with doping, in analogy with 
the Rashba model \cite{sinova2003}. In fact, the
subsequent calculation reveals that the spin Hall conductivity in
this model is nonzero even without doping, which is another
realization of the ``spin Hall insulator''. To calculate the
spin Hall conductivity, we construct an
effective tight-binding model, which reduces to
Eq.~(\ref{L-point}) near the $L$ points. The simplest tight-binding
model is given by the Hamiltonian Eq.~(\ref{Hamiltonian}) where
$ \Gamma_{i}=\tau_{2}\sigma_{i}\  (i=1,2,3)$, $
\Gamma_{4}=\tau_{1}$, $\Gamma_{5}=\tau_{3}$, and
\begin{eqnarray*}
&&d_{1}=C_{1}
\left[\sin{\theta}_{2}-
\sin{\theta}_{3}\right.
+\left. \sin(\theta_{1}-\theta_{2})
-\sin(\theta_{1}-\theta_{3})\right],\\
&&d_{2}=C_{1}
\left[\sin\theta_{3}-
\sin{\theta}_{1}\right.
+\left. \sin(\theta_{2}-\theta_{3})
-\sin(\theta_{2}-\theta_{1})\right],\\
&&d_{3}=C_{1}
\left[\sin\theta_{1}-
\sin\theta_{2}\right.
+\left. \sin(\theta_{3}-\theta_{1})
-\sin(\theta_{3}-\theta_{2})\right],\\
&&d_{4}=C_{2}
\left(
\sin(\theta_{2}+\theta_{3})-
\sin(2\theta_{3}-\theta_{1})\right.
\\
&&\ \ \ \ \ \ \ - \sin(2{\theta}_{2}-{\theta}_{1})+
\sin({\theta}_{2}+{\theta}_{3}-2{\theta}_{1})\\
&&\ \ \ +(\mbox{two cyclic permutations of the subscripts 1,2,3}),\\
&&d_{5}=Mv^{2},
\end{eqnarray*}
where $C_{1}=\frac{\lambda v}{\sqrt{3}a}$ and  
$C_{2}=\frac{v}{4\sqrt{3}a}$. 
It contains a nearest-neighbor and a third-neighbor hopping. Let
us calculate the spin Hall conductivity corresponding to a spin
current $J_{i}^{l}= \frac{1}{2}\left\{J_{i},\sigma_{l}\right\}$ in
response to an external electric field $E_{j}$. Henceforth, 
the Greek indices run from 1 to 3, while the Roman indices run
from 1 to 5. From the Kubo formula, we obtain the result
\begin{eqnarray*}
&&\sigma_{ij}^{l}= \frac{1}{V}\sum_{\mathbf{k}}
\frac{n_{F}(\epsilon_{-})-n_{F}(\epsilon_{+})}{2d^{3}}
\nonumber  \\
&&\ \ \cdot\left( \epsilon_{\alpha m n\beta\gamma}\epsilon_{lmn}
\frac{\partial d_{\alpha}}{\partial k_{i}} \frac{\partial
d_{\gamma}}{\partial k_{j}}d_{\beta}
+2\epsilon_{lmn}\frac{\partial \epsilon}{\partial k_{i}}
\frac{\partial d_{n}}{\partial k_{j}}d_{m}\right)
\end{eqnarray*}
where 
$n_{F}(\epsilon)=(1+e^{\beta(\epsilon-\mu)})^{-1}$.
When we consider again the conserved spin current,
we get
\[
\sigma_{ij(c)}^{l}= \frac{1}{V}\sum_{\mathbf{k}}
\frac{n_{F}(\epsilon_{-})-n_{F}(\epsilon_{+})}{2d^{3}}
\epsilon_{\alpha m n\beta\gamma}\epsilon_{lmn} \frac{\partial
d_{\alpha}}{\partial k_{i}} \frac{\partial d_{\gamma}}{\partial
k_{j}}d_{\beta}.\]
It no longer
depends on $\epsilon(\mathbf{k})$. For the gapped case with only
the lower band is occupied, the spin Hall conductivity
$\sigma_{s}$ is plotted for various values of 
two dimensionless parameters $Mva$ and $\lambda$
in Fig.\ \ref{fig-narrow}. It is an odd function of both $Mva$ and $\lambda$.
Since the Hamiltonian Eq.~(\ref{L-point}) has
eigenenergies
\begin{equation}
\epsilon_{\pm}
=\pm\sqrt{v^{2}(\mathbf{k}\cdot\hat{p})^{2}+\lambda^{2}v^{2}(\mathbf{k}\times
\hat{p})^{2}+(Mv^{2})^{2}}
\end{equation}
near each $L$ point, the values of the parameters can be extracted
from experimental data; $2Mv^{2}$ is a direct gap at the $L$
points, $M$ is an effective mass along the $\hat{p}$ direction, and
$M/\lambda^{2}$ is an effective mass perpendicular to the $\hat{p}$
direction.
In PbS, PbSe and PbTe, the values of the parameters are given 
as $\lambda\sim 1.2, 1.4, 3.3$, and 
$Mva\sim 0.26, 0.16, 0.35$, respectively. The nominal values of the spin Hall
conductivity for these compounds is around $-0.04e/a$, as seen from 
Fig.\ \ref{fig-narrow}.

{\it Concluding remarks.}--- We theoretically predict the spin
Hall effect in zero-gap semiconductors like HgTe and in narrow-gap
semiconductors like PbTe. From a simple tight-binding model, the
spin Hall conductivity is estimated to be of the order of $e/a$,
where $a$ is a lattice constant.
This effect is protected by the band gap $E_G$, which is of
the order of 0.15-0.3eV for narrow-gap semiconductors, 
and hence is robust against the
thermal agitations, impurity scatterings, and electron-electron
inelastic scatterings as long as the energy scale of these is
smaller than $E_G$. In contrast to the doped semiconductors, the
dissipationless spin current in band insulators does not lead to
spin accumulation at the boundary, because it lacks any
mechanisms which breaks time reversal symmetry. Nonetheless, it
can be detected by the electric field
due to the Aharonov-Casher effect 
\cite{meier2003}, which propagates for a 
macroscopic distance via the spin current.

We thank D.~Culcer, E.~Fradkin, A.~H.~MacDonald, Q.~Niu, 
J.~Sinova for helpful discussions. This work is supported by
NAREGI and 
Grant-in-Aids for Scientific Research 
from 
the Ministry of Education, Culture, Sports,
Science and Technology of Japan, the US NSF under grant number
DMR-0342832, and the US Department of Energy, Office of Basic
Energy Sciences under contract DE-AC03-76SF00515.


\begin{thebibliography}{99}
\bibitem{note-nLnH}
We take the hole picture where only the LH is occupied.
\end{thebibliography}
\end{document}